# Improving occupant comfort through real-time Predictive control of Indoor environment using Predicted-Mean Vote model and MLP

Madhan Kumar S[1], Yaswanth Kannan G[2],Kavin Krishna K[3] ,Berlin Hency V[4]

[1]UG Student, B. Tech (ECE), School of Electronics Engineering, VIT Chennai

Madhankumar.2020@vitstudent.ac.in

[2]UG Student, B. Tech (ECE), School of Electronics Engineering, VIT Chennai

yaswanthkannan.g2020@vitstudent.ac.in

[3]UG Student, B. Tech (ECE), School of Electronics Engineering, VIT Chennai

kavinkrishna.kp2020@vitstudent.ac.in

[4]Professor, School of Electronics Engineering, VIT Chennai

berlinhency.victor@vit.ac.in

*Abstract*— **The Predicted Mean Vote (PMV) index is a widely accepted method in the building automation sector because it can precisely estimate indoor thermal comfort levels depending on a variety of environmental parameters. This study suggests an experimental setup for automated real-time optimization of heating, ventilation and air-conditioning (HVAC) operations in closed spaces utilizing PMV-based modelling and Multilayer perceptron (MLP) based prediction, with an experimental setup which includes ESP32 and BME280 sensor. The main objective of this paper is to employ the MLP algorithm and predicted mean vote model to dynamically predict comfort differences considering the fluctuations in environmental conditions such as temperature and humidity. The proposed method is implemented across various settings, encompassing both an anechoic chamber and laboratory environments. The findings indicate that the proposed approach effectively enhances the management of HVAC systems in confined areas, leading to higher energy efficiency and enhanced indoor thermal comfort. The effectiveness of the suggested solution is supported by experimental validation, which shows that significant energy savings are achieved while maintaining comfort levels for occupants within.**

I. Introduction

Predicted-Mean Vote (PMV) plays a crucial role in enhancing building occupant comfort, productivity, and well-being, leading to the increased adoption of building automation and control systems to improve indoor environmental quality [2]. Real-time predictive control algorithms have emerged as a popular choice due to their ability to ensure precise and reliable control of indoor environments, optimizing occupant comfort while maintaining energy efficiency. The PMV model, widely employed for building thermal comfort assessments, utilizes environmental parameters such as temperature, humidity, air velocity, and metabolic rate to predict occupants' mean thermal sensation [1,2]. This study proposes leveraging the PMV model as the basis for a predictive control algorithm aimed at further enhancing occupant comfort.

The MLP employs a deep learning model to predict occupant comfort in real time. To learn the complex nonlinear relationships between environmental parameters and the PMV model, the MLP model is trained using historical data from sensors. The MLP model's output is then used to optimize occupant comfort by adjusting the environmental entities.

To summarize, the paper contains four sections, namely:

- **Literature review**: Investigation of predictive control techniques for HVAC systems and estimation of Predicted-Mean Vote (PMV) index.
- **Proposed methodology**: Integration of algorithms and models to precisely determine occupant comfort, including calibration and testing of the system.
- **Results and discussion**: Presentation of experimental findings, including a comparative analysis with established frameworks.
- **Conclusion**: Emphasis on the significance of the proposed system in improving occupant comfort and well-being, and its potential to optimize closed environments for enhanced comfort.

II. Literature Review

The literature review spotlights the endeavours of a multitude of researchers in the exploration of predictive control methods for HVAC systems and the estimation of Predictive-Mean Vote Index. Hamza Zahid , Oussama Elmansoury , Reda Yaagoubi (2021), proposed a dynamic approach for estimating predicted mean vote [1], which gave insight about the overall factors and BIM modelling of indoor environment. P.O. Fanger (1970), proposed a quantitative study on thermal comfort [2], which focused on deriving a mathematical relation between comfort and environmental parameters. Similarly, ASHRAE Standard 55-2020 [3], highlights the significance of maintaining appropriate thermal conditions in indoor spaces to ensure occupant comfort and well-being. Qiantao Zhao, Zhiwei Lian, Dayi Lai (2021), proposed a review on thermal comfort models and their developments [4], which suggests some future developing directions of thermal comfort models. Faridah Hani Mohamed Salleh, Mulyana binti Saripuddin, Ridha bin Omar (2020), proposed a study on the prediction of thermal comfort in HVAC buildings by considering six thermal factors [5], which outlines the usage of 6 overall factors to predict PMV. Jiawei Yang, Huamin Chen, Shaofu Lin, Limin Chen, Yu Chen (2022), proposed is a study that utilizes a CNN-LSTM model to predict changes in temperature [6], taking into account various environmental factors with multiple dimensions. Tanaya Chaudhuri, Deqing Zhai, Yeng Chai Soh, Hua Li, Lihua Xie and Xianhua Ou (2018), proposed a study that utilized wearable technology to detect the thermal state of occupants [7], with an emphasis on customizing indoor environments based on individual preferences. Ivars Beinarts (2013), proposed a study on optimizing passenger comfort in moving vehicles [8], which utilized a fuzzy logic control method for HVAC equipment.

In this research paper, a novel approach called "InoVAC" is proposed to predict occupant comfort by utilizing ESP32 and edge computing techniques. InoVAC involves the utilization of a plastic container as a replica of a closed environment, where a fan for ventilation, a heating device, and a humidifier for cooling are integrated. Fanger's thermal comfort equation is employed to estimate the Predicted Mean Vote (PMV) value and forecast the level of comfort experienced by the occupant [2]. To enhance the accuracy of the thermal comfort model, a deep learning framework is introduced, which analyzes variations in temperature and humidity within the enclosed environment to forecast the PMV. Through the utilization of IoT devices and edge computing, the approach aims to automate environments, resulting in improved occupant comfort. InoVAC can be implemented in various settings such as commercial establishments, healthcare facilities, and laboratories, reducing the need for extensive human resources. The research signifies a significant advancement in occupant comfort prediction, revolutionizing the field and offering a path to intelligently optimize occupants' well-being in the future.

### III. Methodology

A. *Proposed methodology:*

InoVAC utilizes a fusion of algorithms and models to precisely ascertain occupant comfort levels. Initially, it computes the PMV index by considering fluctuating temperature and humidity parameters within the setting. Subsequently, it categorizes the comfort level according to the PMV index. Lastly, it introduces a standardized algorithm for automating the environmental conditions.

In the research conducted, the widely recognized and ASHRAE-endorsed Fanger's Thermal Comfort Model [2] is employed to estimate the Predicted-Mean Vote (PMV) index [3], taking into account various environmental factors including indoor air temperature, relative humidity, average radiant temperature, water vapor pressure, air velocity, as well as human variables such as metabolic rate and clothing insulation. To enhance occupant comfort, a convolutional neural network (CNN) is utilized [15], which utilizes temperature and humidity as inputs to predict the PMV index. Through meticulous training on a large dataset, the CNN demonstrates remarkable proficiency in accurately forecasting the PMV index across diverse environmental conditions. The integration of the CNN into the thermal comfort model contributes to a deeper understanding and improved performance of the system.

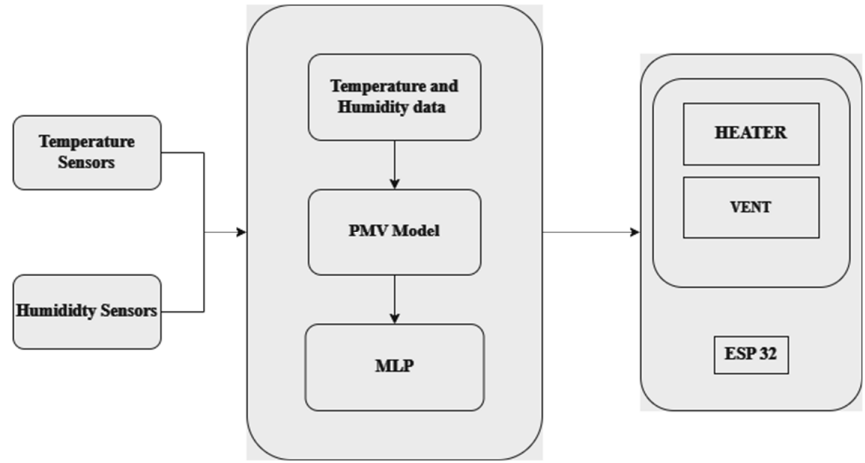

Fig. 1. InoVAC methodology

B. *Clothing insulation and water vapor pressure estimation*

In Algorithm 1, the function *findTCL*, calculates the clothing surface temperature based on the metabolic rate, effective mechanical work done, mean radiant temperature and clothing insulation.

**ALGORITHM 1:**

Clothing surface temperature estimation with Newton-Raphson method.

1) *Define function "**findTCL**" which takes mean radiant temperature, metabolic rate, effective mechanical workdone, and clothing insulation.*
2) *Define a nested **function(tcl)** to calculate clothing surface temperature*
3) *Initialize tcl with the initial tcl0 .*
4) *Iterate for i = 1 to 100:*
   a) *Apply the Newton-Raphson method to find the new value of tcl*
      *Initialize tclNew = Newton_raphson(func,tcl);*
   b) *Check convergence criteria:*
      *If |tclNew - tcl| < 0.00001 and tclNew > 0*
         *Proceed to step4*
      *Else*
         *Return 25 as a default value indicating failure to converge.*
   c) *Update tcl with tclNew*
5) *Return tclNew.*
6) *If the maximum number of iteration failure to converge, return None to indicate failure.*

*End of Algorithm*

The algorithm is based on the Newton-Raphson method to determine the clothing surface temperature (tcl) with a given equation for a specified tolerance of 0.0001 and initial temperature 25ºC.

$$t_{cl} = 35.7 - 0.028 * (M - W) - I_{cl} * \{3.96 * 10^{-8} * f_{cl} * [(t_r+273)^4 - (t_{cl}+273)^4] + f_{cl} * h_c * (t_{cl} - t_a)\} \quad (1)$$

Equation (1) above represents the clothing surface temperature calculation for predicting PMV index, where "M" represents metabolic rate, "W" represents effective work done, "$I_{cl}$" represents clothing insulation, "$f_{cl}$" represents clothing condition, "$h_c$" represents convection heat transfer, "$t_r$" represents mean radiant temperature and "$t_a$" represents air temperature.

Heat transfer equation is applied to calculate the convection heat transfer coefficient from cloth to skin of the occupant

$$h_c = \begin{cases} 2.38 * (t_{cl} * t_a)^{0.25} & for \ 2.38 * |t_{cl} * t_a|^{0.25} \geq 12.1 * \sqrt{va} \\ 12.1 * \sqrt{va} & for \ \ 2.38 * |t_{cl} * t_a|^{0.25} \geq 12.1 * \sqrt{va} \end{cases} \quad (2)$$

Equation (2) is dependent on some environmental parameters such as, "$va$" which represents air velocity of the closed environment.

**ALGORITHM 2:**

Water vapour pressure measurement with Magnus-Tetens equation.

1) *Define the function "findPressure" which Magnus-Tetens equation.*
2) *Compute pressure using the **Magnus – Tetens** equation.*
3) *Return Vapour Pressure.*

*End of algorithm.*

The algorithm is based on Magnus-Tetens equation to determine the atmospheric vapour pressure ($p_a$).

$$P_a = R_h * \left(610.6 * e^{\frac{17.260 * T_a}{273.3 + T_a}}\right) \quad (3)$$

Equation (3) is dependent on environmental parameters such as, "$R_h$" which represents relative humidity and "$T_a$" which represents air temperature.

Following the incorporation of temperature and humidity values into the aforementioned algorithms to derive the clothing surface temperature and water vapor pressure, the computation of the Predicted-Mean Vote (PMV) index is subsequently embarked upon. This pivotal index serves as a comprehensive metric for evaluating thermal comfort.

$$PMV = (0.303 * e^{-0.0036 * M} + 0.028) * \{ M - 3.05 * 10^{-3} * (5733 - 6.99 * M - P_a) - 0.42 * (M - 58.15) - 1.7 * 10^{-5} * M * (5867 - P_a) - 0.0014 * M * (34 - T_a) - 3.96 * 10^{-8} * f_{cl} * [(t_{cl} + 273)^4 - (t_r + 273)^4] - f_{cl} * h_c * (t_{cl} - t_a) \} \quad (4)$$

Equation (4) illustrates the relationship between the predicted mean vote index and various environmental and human factors [1,2], such as indoor air temperature ($T_a$), air velocity ($V_a$), mean radiant temperature($t_r$), relative humidity ($R_h$), metabolic rate (M), clothing condition($f_{cl}$), clothing surface temperature ($t_{cl}$), convection heat coefficient ($h_c$) and effective mechanical power(W).

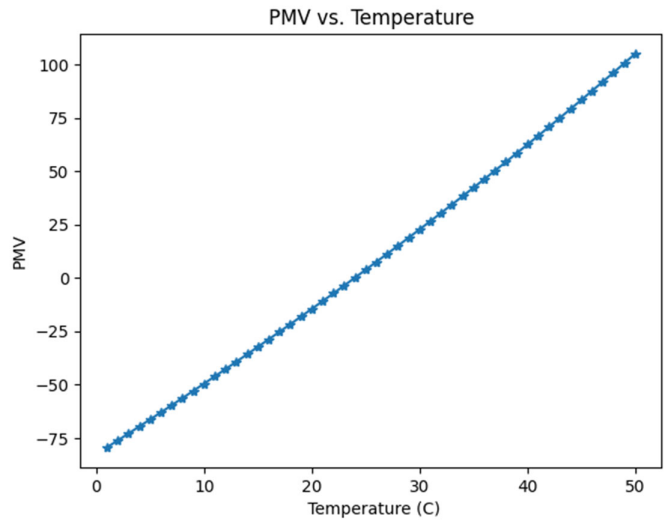

Fig 2. PMV vs Temperature

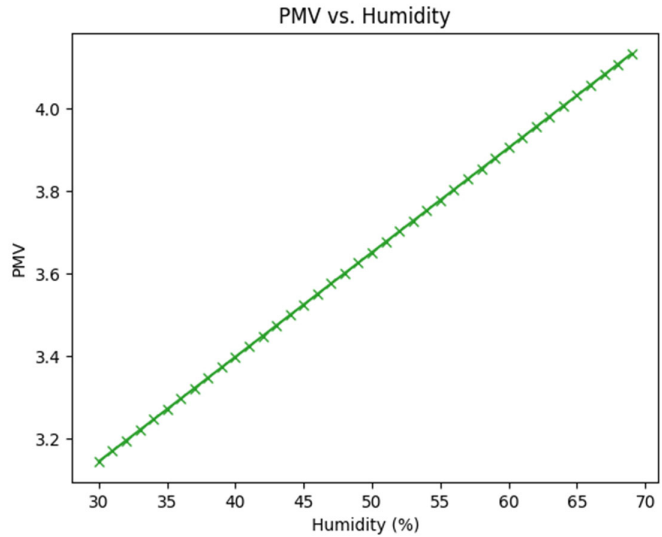

Fig 3. PMV vs Humidity

Two compelling graphical representations have been yielded by the comprehensive investigation, illuminating the intricate relationship between the Predicted-Mean Vote (PMV) index and crucial environmental parameters, namely temperature and humidity. In Fig 2, a captivating trend is showcased where the PMV index demonstrates an exponential ascent, coinciding with escalating temperatures within the enclosed environment. Furthermore, Fig 3 exhibits a linear progression, illustrating the direct impact of increasing humidity on the PMV index. The plots offer profound intuitions into the interplay between environmental variables and the estimation of the predicted-mean vote, employing the thermal comfort mode [2]. This research offers a significant contribution to the field, serving as a valuable resource for the exploration of the intricate dynamics between environmental parameters and human thermal comfort, which is highly sought after.

C. *Importance of Dynamic and Real-time PMV*

As referenced in [3], the predicted-mean vote is notably influenced by ambient temperature and humidity. Numerous scholarly investigations have endeavored to generate a real-time and responsive PMV model using data related to temperature and humidity [1]. By employing the calculated PMV value, it becomes feasible to evaluate indoor thermal comfort effectively. These endeavors not only contribute to the advancement of understanding in the field but also offer a valuable framework for assessing and enhancing the comfort levels experienced in indoor environments.

**ALGORITHM 3:**

Algorithm to classify the comfort levels in enclosed environment.

1) If -0.5 < PMV < 0.5:
    *"COMFORTABLE"*
2) Else if PMV > +2:
    *"HOT"*
    *activateExhaust()*
    *activateCoolants()*
3) Else if PMV < -2:
    *"COLD"*
    *activateHeaters()*

*End of Algorithm*

Based on ISO 7730 guidelines, the environment is said to have a "best thermal comfort" when the PMV index falls within the range of -0.5 to +0.5. If the PMV index ranges from +0.5 to +2, the environment is perceived as "warm," while exceeding +2 categorizes it as "hot". Conversely, falling within the range of -0.5 to -2 indicates a "cool" environment, while a PMV index below -2 denotes a "cold" environment.

### D. *Predicted-mean vote as a function of temperature and humidity*

The study presents an innovative method that leverages a convolutional neural network (CNN) to predict the PMV index [3] using temperature and humidity as input parameters. By improving the precision of the thermal comfort model, this method has the capacity to make a valuable contribution to energy conservation initiatives.

Table 1: Specifics of Custom Dataset for training the Network

| S. No. | Specifics | Figures |
|---|---|---|
| 1. | Entities | 50000 |
| 2. | Temperature | 0 °C - 50 °C |
| 3. | Humidity | 0% - 100% |

The convolutional neural network was meticulously trained on a comprehensive dataset, showcasing outstanding precision in predicting the PMV index under diverse environmental circumstances. This approach presents a pragmatic resolution to tackle issues linked to thermal comfort model variations due to environmental changes, ultimately resulting in a significant enhancement in occupant comfort.

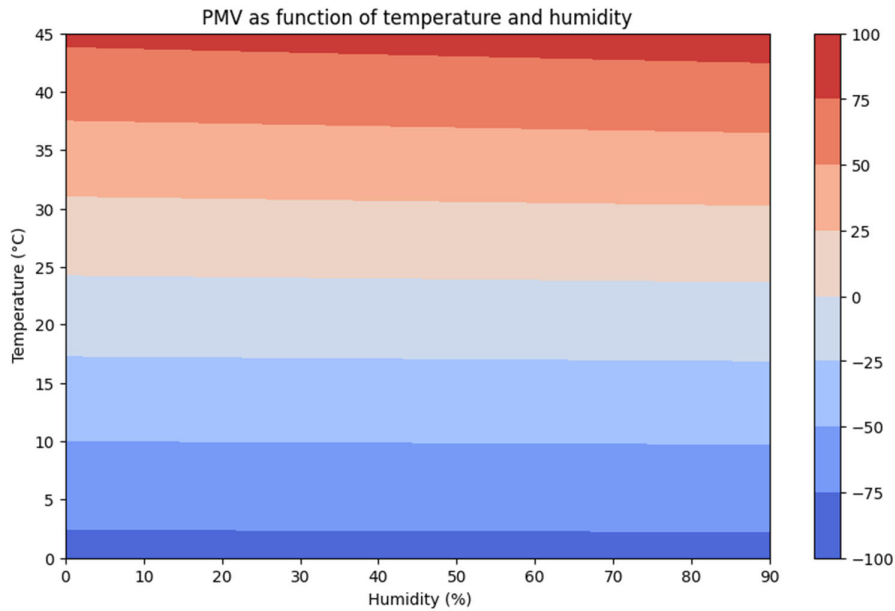

Fig 4. Influence of Temperature and Humidity on PMV

The intricate convolutional neural network (CNN) [15, 18] put forward consists of a total of six layers, including an initial input layer, four hidden layers, and a final output layer. Within the model, the dense layers incorporate the ReLU activation function to introduce non-linearity and capture complex patterns. The final output layer employs a sigmoid activation function for easier classification. The model is compiled with the mean squared error loss function and trained using the Adam optimizer. Notably, the PMV index is predicted using temperature and humidity as input variables.

Table 2: InoVAC-CNN Model Performance

| S. No. | Metrics | Value |
|---|---|---|
| 1. | Mean Squared Error (MSE) | 0.004035 |
| 2. | Mean Absolute Error (MAE) | 0.01318 |
| 3. | R-Squared Score ($R^2$) | 0.985653 |

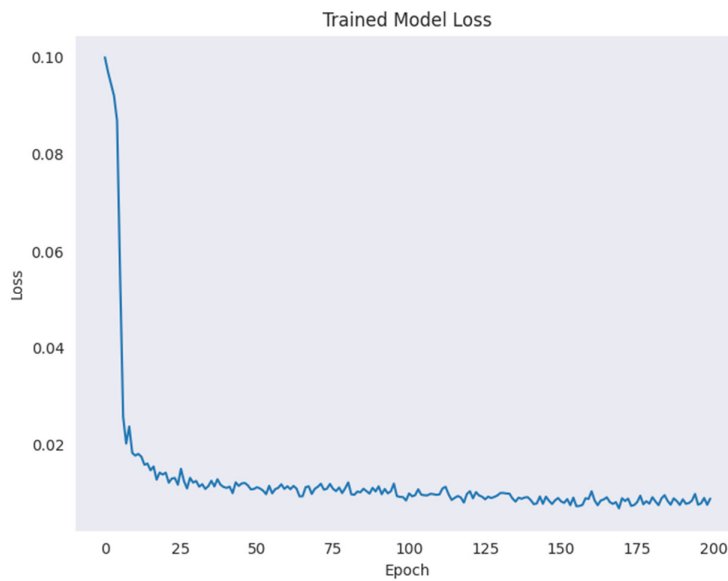

Fig 5. Model's Training-Loss Graph

The performance of the InoVAC-CNN model was evaluated using several metrics, including mean squared error (MSE), mean absolute error (MAE), and the R-squared score. It demonstrated remarkable precision, with minimal predictive discrepancies: an MSE of 0.004035 and an MAE of 0.01318. Furthermore, a substantial R-squared score of 0.985653 highlighted the model's robust capacity to explain the variability in the data.

Figure 6 illustrates that the InoVAC-CNN model successfully forecasts comfort levels through the use of temperature and humidity as input variables, ultimately improving the comfort experienced by occupants.

```
Enter the temperature (in Celsius): 23.45
Enter the humidity (in percentage): 45.67
1/1 [==============================] - 0s 96ms/step
Predicted PMV: 0.256
```

Fig 6. Forecasting PMV using InoVAC-CNN.

Following the training phase, the essential elements of the InoVAC-CNN model, comprising weights, biases, and structure, were extracted. Subsequently, an efficient Multi-Layer Perceptron (MLP) was created to perform inference on an ESP32 device [17]. The projected "PMV" from the CNN model was integrated into the algorithm, as detailed in section III-C. This approach simplifies the process of implementing the InoVAC model on a low-power embedded device, allowing for real-time inference even in resource-constrained environments.

IV. **Experiments and results**

To assess InoVAC's effectiveness, two experiments were conducted. The first experiment involved calibrating the system within an industry-standard anechoic chamber [18], depicted in Figure 7.

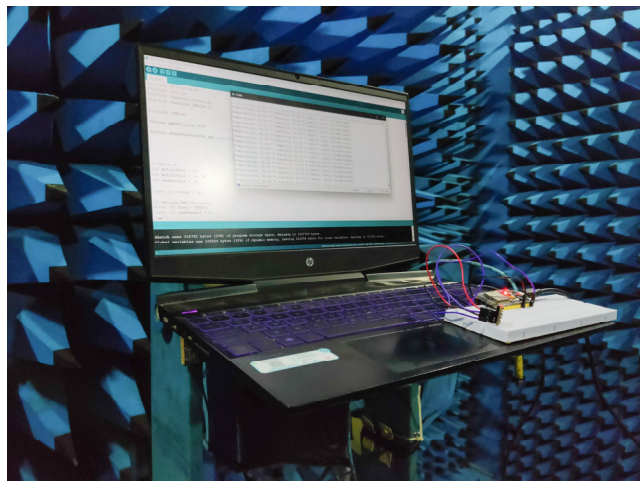

Fig 7. ESP32-BME280 Calibration in the Anechoic Chamber

In the initial experiment, the InoVAC system underwent calibration to establish several parameters, including air temperature and humidity. The chamber's temperature was deliberately altered, allowing for the assessment of the initial temperature within the chamber, as detailed in section III-B.

Table 3: Evaluation of the InoVAC system

| S.No. | Temperature (ºC) | Humidity (%) | Comfort Level |
|---|---|---|---|
| 1. | 32.34 | 62.22 | Uncomfortable |
| 2. | 25.00 | 60.99 | Comfortable |

The calibration results of the InoVAC system, including observed temperature, humidity, and corresponding estimated comfort levels, are presented in Table 3. The calibration process involved two different temperatures: 25 ºC and 32.34 ºC, with humidity levels of 60.99% and 62.22%, respectively. Following the validation process, the initial temperature was set to 25 ºC, as mentioned in section III-B. Anechoic chambers are well-known for providing a controlled environment with constant temperature and humidity, ensuring precise calibration of the device and reliable measurements.

In the second experiment, the InoVAC system was tested in a laboratory setting without ventilation or HVAC devices to evaluate its dependability and efficiency. The system operated as anticipated, accurately predicting a higher PMV value indicative of discomfort. The BME280 sensor provided precise temperature and humidity measurements, which were used as inputs for the MLP to accurately predict the PMV. The outcomes of both experiments affirm the viability and efficiency of the InoVAC system in real-life situations.

Table 4: Comparison with currently established methods.

| S.No. | Methodoolgy | Impact Factors | Accuracy |
|---|---|---|---|
| 1. | InoVAC | • Use of 5 environment variable and 6 human variables<br>• High PMV indoor environment<br>• Well calibrated thermal comfort model along with automation of HVAC elements | 99.6% |
| 2 | [5] | • 6 Factors are considered<br>• Comparison among ML models<br>• Commercial buildings | 80% |
| 3 | [6] | • 5 factors are considered<br>• Temperature change prediction<br>• Generic indoor environment dataset | >80% |
| 4 | [7] | • Use of psychological factors and skin temperature<br>• SVM based image classification<br>• Focused for wearable technology | 90.6% |
| 5 | [8] | • Fuzzy logic control<br>• Electricity and voting based analysis<br>• Mobile environments like train | NA |

The InoVAC system distinguishes itself through its comprehensive approach, considering five environmental variables and six human variables, while operating in a high Predicted Mean Vote (PMV) indoor environment. It incorporates a meticulously calibrated thermal comfort model and automation of HVAC elements, resulting in an impressive accuracy of 99.6 In comparison, existing systems have limitations. Some systems only consider six overall factors in commercial buildings, with an accuracy of 80%. Similarly, systems focusing on five overall factors in generic indoor environments achieve accuracy above 80%. Other systems concentrate on psychological factors and skin temperature using wearable technology, reaching an accuracy of 90.6%. Fuzzy logic control combined with electricity and voting-based analysis is utilized in mobile environments like trains. These comparisons highlight the unique characteristics and superior accuracy of the InoVAC system in ensuring indoor occupant comfort. It finds effective applications in sustainable environments such as smart grid systems and energy-efficient retrofitting.

## V. Conclusion

In conclusion, the InoVAC method has been introduced for indoor HVAC automation, aimed at enhancing occupant comfort by predicting the Dynamic Predicted Mean Vote (PMV) and estimating the corresponding comfort levels based on a thermal comfort model. The InoVAC approach incorporates unique algorithms and models that consider environmental dependencies and human variable dependencies, specifically the variations in temperature and humidity within both poorly ventilated and highly enclosed indoor environments.

To achieve accurate PMV predictions, the InoVAC system has been employed, which includes a feed-forward deep learning model that effectively captures the relationship between temperature, humidity, and thermal comfort. Through comprehensive experiments conducted in standard closed environments, such as an anechoic chamber, as well as in relatively least ventilated high PMV environments like laboratories and manufacturing units, the reliability and effectiveness of the InoVAC system have been demonstrated. Furthermore, precise temperature and pressure measurements are achieved by utilizing an MLP and integrating an ESP32 central device connected to a BME280 sensor, thereby further improving the reliability and accuracy of the indoor system.

By accurately estimating and maintaining occupant comfort in enclosed environments, the InoVAC method contributes to the overall well-being and satisfaction of individuals, offering opportunities for further enhancements and applications.

**Acknowledgment**

The researchers wish to express their heartfelt gratitude to the "Microwave and Antenna Measurement Laboratory" at VIT Chennai for their kind support in granting access to their state-of-the-art anechoic chamber, which was instrumental in evaluating the system.

**References**

[1] Hamza Zahid, Oussama Elmansoury, Reda Yaagoubi, "Dynamic Predicted Mean Vote: An IoT-BIM integrated approach for indoor thermal comfort optimization". *Automation in Construction 129 (2021) 103805*.

[2] P.O. Fanger, Thermal Comfort: Analysis and Applications in Environmental Engineering, Danish Technical Press, Copenhagen, Denmark, 1970, p. 244. ISBN 8757103410, 9788757103410.

[3] American Society of Heating, Refrigerating and Air-Conditioning Engineers (ASHRAE), ANSI/ASHRAE Standard 55-2020 - Thermal Environmental Conditions for Human Occupancy, ASHRAE, Georgia, USA, ISSN 1041–2336.

[4] Qiantao Zhao, Zhiwei Lian, Dayi Lai "Thermal comfort models and their developments: A review". *Energy and Built Environment (2021)*.

[5] Faridah Hani Mohamed Salleh, Mulyana binti Saripuddin, Ridha bin Omar, "Predicting Thermal Comfort of HVAC Building Using 6 Thermal Factors". *2020 8th International Conference on Information Technology and Multimedia (ICIMU)*.

[6] Jiawei Yang, Huamin Chen, Shaofu Lin, Limin Chen, Yu Chen, "Prediction of temperature change with multidimensional environmental characteristic based on CNN-LSTM-ATTENTION model". *2022 IEEE 10th Joint International Information Technology and Artificial Intelligence Conference (ITAIC)*.

[7] Tanaya Chaudhuri, Deqing Zhai, Yeng Chai Soh, Hua Li, Lihua Xie and Xianhua Ou, "Convolutional Neural Network and Kernel Methods for Occupant Thermal State Detection using Wearable Technology". *2018 International Joint Conference on Neural Networks (IJCNN)*.

[8] Ivars Beinarts, "Fuzzy Logic Control Method of HVAC Equipment for Optimization of Passengers' Thermal Comfort in Public Electric Transport Vehicles". *EuroCon 2013*, IEEE, 2013.

[9] Ana Rita Santiago, Mario Antunes, Joao Paulo Barraca, Diogo Gomes and Rui L. Aguiar, "Predictive Maintenance System for efficiency improvement of heating equipment". *2019 IEEE Fifth International Conference on Big Data Computing Service and Applications*. IEEE, 2019.

[10] Fabliha Bushra Islam, Cosmas Ifeanyi Nwakanma, Dong-Seong Kim and Jae-Min Lee, "IoT-Based HVAC Monitoring System for Smart Factory". IEEE, 2020.

[11] Wahiba Yaïci, Evgueniy Entchev and Michela Longo, "Internet of Things (IoT)-Based System for Smart Home Heating and Cooling Control", *2022 IEEE International conference on environment and electrical engineering and IEEE industrial and Commercial Power Systems Europe*. IEEE, 2022.

[12] Kumar Padmanabh, Ahmad Al-Rubaie and Alia Abdulaziz Ali Abdulla Aljasmi, "Health Estimation and Fault Prediction of the Sensors of a HVAC System", *2022 IEEE International Conference on Internet of Things and Intelligence Systems (IoTaIS)*. IEEE, 2022.

[13] Vinay Kumar, Rakesh Kumar, Deepraj Patkar, Ajit S. Bopardikar, "A Method to Identify Dynamic Zones for Efficient Control of HVAC Systems". IEEE, 2017.

[14] Niima Es-sakalia, Moha Cherkaoui, Mohamed Oualid Mghazli and Zakaria Naimi, "Review of predictive maintenance algorithms applied to HVAC systems", *Energy Reports 8 (2022) 1003–1012*.

[15] Larisa Lyutikova, "Correction of Neural Network Solutions by Logical Analysis Methods", International Scientific Conference "INTERAGROMASH 2022".

[16] Abhishek Sebastian, Pragna R, Madhan Kumar S, Jesher Joshua M, Sarath Prathap, Brintha Therese, "Proximity-Based Access Control with BLE Communication Using Path Loss Model and MLP Prediction," *2023 2nd International Conference on Vision Towards Emerging Trends in Communication and Networking Technologies (ViTECoN), Vellore, India, 2023.*

[17] Immonen, R., & Hämäläinen, T. (2022). Tiny Machine Learning for Resource-Constrained Microcontrollers. Journal of Sensors, 2022(2022).

[18] Xu, Qian, and Yi Huang. "Anechoic and Reverberation Chambers: Theory, Design, and Measurements." (2019).